\documentclass[12pt]{amsart}
\usepackage{amssymb,latexsym,amsmath,amscd,graphicx,url,color}
\theoremstyle{plain}

\newtheorem{thm}{Theorem}[section]
\newtheorem{prop}[thm]{Proposition}
\newtheorem{cor}[thm]{Corollary}

\theoremstyle{definition}
\newtheorem{defn}[thm]{Definition}
\newtheorem{rmk}[thm]{Remark}

\newtheorem{notat}[thm]{Notation}

\newtheorem{ex}[thm]{Example}

\newcommand{\cA}{{\mathcal A}}

\newcommand{\C}{{\mathcal C}}
\newcommand{\D}{{\mathcal D}}
\newcommand{\FF}{{\mathbb F}}

\newcommand{\QQ}{{\mathbb Q}}

\newcommand{\ZZ}{{\mathbb Z}}
\newcommand{\cc}{\textnormal{c}}
\newcommand{\rr}{\textnormal{r}}

\DeclareMathOperator{\supp}{supp}

\DeclareMathOperator{\mat}{Mat}
\DeclareMathOperator{\tr}{tr}
\DeclareMathOperator{\rk}{rk}
\DeclareMathOperator{\gln}{GL_n}
\DeclareMathOperator{\glm}{GL_m}
\DeclareMathOperator{\colsp}{colsp}
\DeclareMathOperator{\rowsp}{rowsp}

\title{Rank-metric codes}

\author{E. Gorla}
\address{Institut de Math\'ematiques, Universit\'e de Neuch\^atel, 
Rue Emile-Argand 11, 2000 Neuch\^atel, Switzerland}  
\email{elisa.gorla@unine.ch}

\thanks{Part of this chapter was written while the author was participating in the Nonlinear Algebra program at ICERM in Fall 2018. The author wishes to thank ICERM and Brown University for an excellent working environment.}

\begin{document}
\maketitle

\section{Definitions, isometries, and equivalence of codes}

Let $q$ be a prime power and let $\FF_q$ denote the finite field with $q$ elements. Let $m,n$ be positive integers and denote by $\mat_{n\times m}(\FF_q)$ the $\FF_q$-vector space of matrices of size $n\times m$ with entries in $\FF_q$. 

In this chapter, we discuss the mathematical foundations of rank-metric codes. We restrict our attention to linear codes. All dimensions are over $\FF_q$, unless otherwise stated. 

\begin{defn}
For a matrix $A\in\mat_{n\times m}(\FF_q)$, we let $\rk(A)$ denote the rank of $A$. 
The function 
$$\begin{array}{ccc} d:\mat_{n\times m}(\FF_q)\times\mat_{n\times m}(\FF_q) & \longrightarrow & \mat_{n\times m}(\FF_q) \\ (A,B) & \longmapsto & \rk(A-B) \end{array}$$ 
is a distance on $\mat_{n\times m}(\FF_q)$, which we call \textbf{rank distance} or simply \textbf{distance}. The rank is the corresponding \textbf{weight function}.

A (\textbf{matrix}) \textbf{rank-metric code} is an $\FF_q$-linear subspace $\C \subseteq \mat_{n\times m}(\FF_q)$. 
\end{defn}

A class of rank-metric codes that has received a lot of attention is that of vector rank-metric codes, introduced independently by Gabidulin and Roth in~\cite{Gabidulin1985} and~\cite{Roth1991}.

\begin{defn}
The $\textbf{rank weight}$ $\rk(v)$ of a vector $v \in \FF_{q^m}^n$ is the dimension of the $\FF_q$-linear space generated by its entries. The function
$$\begin{array}{ccc} d:\FF_{q^m}^n\times\FF_{q^m}^n & \longrightarrow & \FF_{q^m}^n \\ (u,v) & \longmapsto & \rk(u-v) \end{array}$$ 
is a distance on $\FF_{q^m}^n$, which we call \textbf{rank distance}  or simply \textbf{distance}. 

A \textbf{vector rank-metric code} is an $\FF_{q^m}$-linear subspace $C \subseteq \FF_{q^m}^n$.  
\end{defn}

Every vector rank-metric code can be regarded as a rank-metric code, up to the choice of a basis of $\FF_{q^m}$ over $\FF_q$.

\begin{defn}\label{defn:lift}
Let $\Gamma=\{\gamma_1,...,\gamma_m\}$ be a basis of $\FF_{q^m}$ over $\FF_q$ and let $v \in \FF_{q^m}^n$. Define $\Gamma(v)\in\mat_{n\times m}(\FF_q)$ via the identity
$$v_i=\sum_{j=1}^m \Gamma_{ij}(v) \gamma_j, \quad i=1,\ldots,n.$$
Let $C \subseteq \FF_{q^m}^n$ be a vector rank-metric code. The set
$$\Gamma(C)=\{\Gamma(v)\mid v\in C\}$$
is the {\bf rank-metric code associated to $C$ with respect to $\Gamma$}.
\end{defn}

\begin{ex}
Let $C$ be the vector rank-metric code $C=\langle(1,\alpha)\rangle\subseteq\FF_8^2$. Let $\FF_8=\FF_2[\alpha]/(\alpha^3+\alpha+1)$ and let $\gamma_1 =1$, $\gamma_2=\alpha$, $\gamma_3=\alpha^2$. Then $\Gamma=\{\gamma_1,\gamma_2,\gamma_3\}$ is a basis of $\FF_8$ over $\FF_2$ and
$$\Gamma(1,\alpha)=\begin{pmatrix} 1 & 0 & 0 \\ 0 & 1 & 0\end{pmatrix},\; 
\Gamma(\alpha,\alpha^2)=\begin{pmatrix} 0 & 1 & 0 \\ 0 & 0 & 1\end{pmatrix},\;
\Gamma(\alpha^2,\alpha+1)=\begin{pmatrix} 0 & 0 & 1 \\ 1 & 1 & 0\end{pmatrix}.$$ Hence
$$\Gamma(C)=\left\langle \begin{pmatrix} 1 & 0 & 0 \\ 0 & 1 & 0\end{pmatrix}, \begin{pmatrix} 0 & 1 & 0 \\ 0 & 0 & 1\end{pmatrix}, 
\begin{pmatrix} 0 & 0 & 1 \\ 1 & 1 & 0\end{pmatrix} \right\rangle\subseteq\mat_{2\times 3}(\FF_2).$$
\end{ex}

The image $\Gamma(C)$ of a vector rank-metric code $C$ via $\Gamma$ as defined above is a rank-metric code, whose parameters are determined by those of $C$. The proof of the next proposition is easy and may be found e.g. in~\cite[Section 1]{Gorla2018ch}.

\begin{prop}\label{prop:lift}
The map $v \mapsto \Gamma(v)$ is an $\FF_q$-linear isometry, i.e., it is a homomorphism of $\FF_q$-vector spaces which preserves the rank.  
In particular, if $C \subseteq \FF_{q^m}^n$ is a vector rank-metric code of dimension $k$ over $\FF_{q^m}$, then $\Gamma(C)$ is an $\FF_q$-linear rank-metric code of dimension $mk$ over $\FF_q$. 
\end{prop}

The following is the natural notion of equivalence for rank-meric codes. 

\begin{defn}\label{defn:equiv}
An {\bf $\FF_q$-linear isometry} $\varphi$ of $\mat_{n\times m}(\FF_q)$ is an $\FF_q$-linear homomorphism $\varphi: \mat_{n\times m}(\FF_q)\to\mat_{n\times m}(\FF_q)$ such that $\rk(\varphi(M))=\rk(M)$ for every $M\in\mat_{n\times m}(\FF_q)$.

Two rank-metric codes $\C,\D \subseteq \mat_{n\times m}(\FF_q)$ are {\bf equivalent} if there is an $\FF_q$-linear isometry $\varphi: \mat_{n\times m}(\FF_q)\to\mat_{n\times m}(\FF_q)$ such that $\varphi(\C)=\D$. If $\C$ and $\D$ are equivalent rank-metric codes, we write $\C\sim\D$.
\end{defn}

Some authors define a notion of equivalence for vector rank-metric codes as follows.

\begin{defn}\label{defn:equiv_vector}
An {\bf $\FF_{q^m}$-linear isometry} $\psi$ of $\FF_{q^m}^n$ is an $\FF_{q^m}$-linear homomorphism $\psi: \FF_{q^m}^n\to\FF_{q^m}^n$ such that $\rk(\psi(v))=\rk(v)$ for every $v\in\FF_{q^m}^n$.

Two vector rank-metric codes $C,D \subseteq \FF_{q^m}^n$ are {\bf equivalent} if there is an $\FF_{q^m}$-linear isometry $\psi: \FF_{q^m}^n\longrightarrow \FF_{q^m}^n$ such that $\psi(C)=D$. If $C$ and $D$ are equivalent vector rank-metric codes, we write $C\sim D$.
\end{defn}

Notice however that Definition~\ref{defn:lift} allows us to apply the notion of equivalence from Definition~\ref{defn:equiv} to vector rank-metric codes. It is therefore natural to ask whether the rank-metric codes associated to equivalent rank-metric codes are also equivalent. It is easy to show that the answer is affirmative.

\begin{prop}[\cite{Gorla2018p}, Proposition 1.15]
Let $C,D \subseteq \FF_{q^m}^n$ be vector rank-metric codes. Let $\Gamma$ and $\Gamma'$ be bases
of $\FF_{q^m}$ over $\FF_q$. If $C\sim D$, then $\Gamma(C)\sim\Gamma'(D)$.
\end{prop}

Linear isometries of $\mat_{n\times m}(\FF_q)$ and of $\FF_{q^m}^n$ can be easily characterized. The following result was shown by Hua for fields of odd characteristic and by Wan for fields of characteristic 2.

\begin{thm}[\cite{Hua1951, Wan1962}]\label{thm:hua}
Let $\varphi:\mat_{n\times m}(\FF_q)\to\mat_{n\times m}(\FF_q)$ be an $\FF_q$-linear isometry with respect to the rank metric. \begin{itemize}
\item If $m\neq n$, then there exist matrices $A \in \gln(\FF_q)$ and $B\in \glm(\FF_q)$ such that $\varphi(M)=AMB$ for all $M \in\mat_{n\times m}(\FF_q)$.
\item If $m=n$, then there exist matrices $A,B \in \gln(\FF_q)$ such that either $\varphi(M)=AMB$ for all $M \in \mat_{n\times n}(\FF_q)$, or $\varphi(M)=AM^\textsf{T}B$ for all $M \in\mat_{n\times n}(\FF_q)$.
\end{itemize}
\end{thm}

The corresponding characterization of isometries of $\FF_{q^m}^n$ was given by Berger.

\begin{thm}[\cite{Berger2002}]\label{thm:berger}
Let $\psi:\FF_{q^m}^n \to \FF_{q^m}^n$ be an $\FF_{q^m}$-linear isometry with respect to the rank metric. 
Then there exist $\alpha\in\FF_{q^m}^*$ and $B\in\gln(\FF_q)$ such that $\psi(v)=\alpha v B$ for all 
$v\in\FF_{q^m}^n$.
\end{thm}

\begin{notat}
For a vector rank-metric code $C\subseteq\FF_{q^m}^n$ and $B\in\gln(\FF_q)$, let
$$CB=\{vB\mid v\in C\} \subseteq \FF_{q^m}^n.$$
For a rank-metric code $\C\subseteq\mat_{n\times m}(\FF_q)$, let $$\C^\textsf{T}=\{M^\textsf{T}\mid M\in\C\}\subseteq\mat_{m\times n}(\FF_q).$$
\end{notat}

The MacWilliams Extension Theorem is a classical result in the theory of linear block codes in $\FF_q^n$ with the Hamming distance. It essentially says that any linear isometry of block codes can be extended to a linear isometry of the ambient space $\FF_q^n$. It is natural to ask whether an analogue of the MacWilliams Extension Theorem holds for rank-metric codes. In other words, given rank-metric codes $\C,\D\subseteq\mat_{n\times m}(\FF_q)$ and an $\FF_q$-linear isometry $f:\C\to\D$, one may ask whether there exists an $\FF_q$-linear isometry $\varphi:\mat_{n\times m}(\FF_q)\to\mat_{n\times m}(\FF_q)$ such that $\varphi\mid_{\C}=f$. The answer is no, as the next example shows. More counterexamples can be found in~\cite{Barra2015} and in the preprint~\cite[Section~7]{delaCruz2015u}.

\begin{ex}[\cite{Barra2015}, Example 2.9 (a)]
Denote by $0$ the zero matrix of size $2\times 1$ and let $$\C=\{(A \ 0)\mid A \in\mat_{2\times 2}(\FF_2)\}\subseteq \mat_{2\times 3}(\FF_2).$$
Let $\varphi:\C\to\mat_{2\times 3}(\FF_2)$ be defined by $\varphi (A\ 0)=(A^\textsf{T}\ 0)$. Then $\varphi$ is an $\FF_2$-linear isometry defined on $\C$ which is not the restriction to $\C$ of an $\FF_2$-linear isometry of $\mat_{2\times 3}(\FF_2)$. In fact, there is no choice for $\varphi\begin{pmatrix} 0 & 0 & 1 \\ 0 & 0 & 0 \end{pmatrix}$ that preserves the property that $\varphi$ is an $\FF_2$-linear isometry.
\end{ex}

\section{The notion of support in the rank-metric}

In analogy with the notion of support of a codeword for linear block codes, one may define the support of a codeword in a vector rank-metric code. For a matrix $M\in\mat_{n\times m}(\FF_q)$, we denote by $\colsp(M)\subseteq\FF_q^n$ the $\FF_q$-vector space generated by the columns of $M$ and by $\rowsp(M)\subseteq\FF_q^m$ the $\FF_q$-vector space generated by the rows of $M$.

\begin{defn}\label{defn:supp_vect}[\cite{Jurrius2017}, Definition~2.1]
Let $C\subseteq\FF_{q^m}^n$ be a vector rank-metric code, and let $\Gamma=\{\gamma_1,\ldots,\gamma_m\}$ be a basis of $\FF_{q^m}$ over $\FF_q$.
The {\bf support} of $v\in C$ is the $\FF_q$-linear space $$\supp(v)=\colsp(\Gamma(v))\subseteq\FF_q^n.$$ The {\bf support} of a subcode $D\subseteq C$ is $$\supp(D)=\sum_{v\in D}\supp(v)\subseteq\FF_q^n.$$
\end{defn}

Notice that $\supp(v)$ does not depend on the choice of the basis $\Gamma$, since if $\Gamma'$ is another basis of $\FF_{q^m}$ over $\FF_q$, then there exists a $B\in\glm(\FF_q)$ such that $\Gamma(v)=\Gamma'(v)B$. This also implies that $\supp(D)$ does not depend on the choice of $\Gamma$. See also~\cite[Proposition~1.13]{Gorla2018p}.

In the context of rank-metric codes, we define the support as follows.

\begin{defn}\label{defn:supp}
Let $\C\subseteq\mat_{n\times m}(\FF_q)$ be a rank-metric code.\\ \noindent
If $n\leq m$ define the {\bf support} of $M\in\C$ as the $\FF_q$-linear space $$\supp(M)=\colsp(M)\subseteq\FF_q^n.$$
If $n>m$ define the {\bf support} of $M\in\C$ as the $\FF_q$-linear space $$\supp(M)=\rowsp(M)\subseteq\FF_q^m.$$
The {\bf support} of a subcode $\D\subseteq\ C$ is $$\supp(\D)=\sum_{M\in \D}\supp(M)\subseteq\FF_q^{\min\{m,n\}}.$$
\end{defn}

Notice that, if $n\leq m$, Definition~\ref{defn:supp} agrees with Definition~\ref{defn:supp_vect}, when restricted to rank-metric codes associated to vector rank-metric codes. Precisely, if $C\subseteq\FF_{q^m}^n$ is a vector rank-metric code and $\Gamma=\{\gamma_1,\ldots,\gamma_m\}$ is a basis of $\FF_{q^m}$ over $\FF_q$, then $$\supp(\Gamma(v))=\supp(v)$$ for all $v\in C$, under the assumption that $n\leq m$. 

\begin{rmk}
If $n>m$, the support of $v\in\FF_{q^m}^n$ according to Definition~\ref{defn:supp_vect} is $\colsp(\Gamma(v))$, while the support of $\Gamma(v)\in\mat_{n\times m}(\FF_q)$ according to Definition~\ref{defn:supp} is $\rowsp(\Gamma(v))$. In other words, Definition~\ref{defn:supp} for the elements of the rank-metric code associated to a vector rank-metric code does not coincide with Definition~\ref{defn:supp_vect} for the elements of the vector rank-metric code. This will not create confusion, since our notation allows us to distinguish the two situations: $\supp(v)=\colsp(\Gamma(v))$ while $\supp(\Gamma(v))=\rowsp(\Gamma(v))$, if $n>m$.
\end{rmk}

We wish to stress that, in the context of matrices, taking the support of the transposed yields a different, but well-behaved notion of support of a matrix. Below we make a few remarks on different possible notions of support and on why we choose to adopt Definition~\ref{defn:supp}. It is clear that, depending on the application or on the information that one wishes to encode, one may also choose to work with different notions of support.

\begin{rmk}\label{rmk:supp_square}\index{support!for rank-metric codes!for square matrices}
If $n=m$, then one may define a notion of support by considering row spaces instead of column spaces. This yields a different, but substantially equivalent notion of support. A different, but possibly interesting, notion of support for a square matrix would be defining the support of $M\in\mat_{n\times n}(\FF_q)$ to be the pair of vector spaces $(\rowsp(M),\colsp(M))$. This is connected to the definition of generalized weights (see Section~\ref{sect:genwgts}) and to the approach taken in~\cite{Gorla2018p} for studying generalized weights of square matrices via $q$-polymatroids (see Section~\ref{qpolym}).
\end{rmk}

\begin{rmk}
For any value of $m,n$, one may define a different notion of support as follows: 
For a rank-metric code $\C\subseteq\mat_{n\times m}(\FF_q)$ and for $M\in\C$, let 
\begin{equation}\label{eqn:supp1}
\supp(M)=\rowsp(M)\subseteq\FF_q^m\;\; \mbox{ if } n\leq m
\end{equation}
and let 
\begin{equation}\label{eqn:supp2}
\supp(M)=\colsp(M)\subseteq\FF_q^n\;\; \mbox{ if } n>m.
\end{equation}
Then the support of a subcode $\D\subseteq\ C$ is 
\begin{equation}\label{eqn:supp3}
\supp(\D)=\sum_{M\in \D}\supp(M)\subseteq\FF_q^{\max\{m,n\}}.
\end{equation}
This definition yields a different notion of support from that of Definition~\ref{defn:supp}, in particular it takes values in $\FF_q^{\max\{n,m\}}$. One can check that both notions of support are regular in the sense of~\cite{Ravagnani2018}. However, the definition of support in~(\ref{eqn:supp1}), (\ref{eqn:supp2}), and~(\ref{eqn:supp3}) for $n\neq m$ yields an empty extremality theory in the sense of~\cite[Section~7]{Ravagnani2018} and a series of redundant MacWilliams Identities. Moreover, in~\cite{Gorla2018p} we showed that, for $n<m$, the $q$-polymatroid determined by the supports as in Definition~\ref{defn:supp} allows one to easily recover the generalized weights of the code, while the $q$-polymatroid determined by the supports as in~(\ref{eqn:supp1}), (\ref{eqn:supp2}), and~(\ref{eqn:supp3}) does not. With these in mind, we choose to adopt Definition~\ref{defn:supp}. 
\end{rmk}

\begin{rmk}\label{rmk:suppcols}
Some authors choose to work with a definition of support which is the same for every value of $m,n$; e.g. in~\cite{MartinezPenas2018} the authors define 
\begin{equation}\label{eqn:supp4}
\supp(M)=\colsp(M)\;\; \mbox{ for any } m,n \mbox{ and  any } M\in\mat_{n\times m}(\FF_q).
\end{equation}
This notion of support agrees with Definition~\ref{defn:supp} for $n\leq m$ and with the definition discussed in the previous remark for $n>m$. For all the reasons discussed in the previous remark, for $n>m$ we prefer Definition~\ref{defn:supp} to this definition. Notice however that the definition of support in~(\ref{eqn:supp4}) is compatible with the notion of generalized matrix weights as defined by Mart{\'\i}nez-Pe\~nas and Matsumoto (see Definition~\ref{defn:higherwtsMP}). In this chapter, however, we define generalized weights as in Definition~\ref{defn:higherwts}, which is compatible with the notion of support as by Definition~\ref{defn:supp}.
\end{rmk}

Given a definition of support, it is natural to consider the subcodes of a code which are supported on a fixed vector space.

\begin{defn}
Let $\C\subseteq\mat_{n\times m}(\FF_q)$ be a rank-metric code. Let $V\subseteq\FF_q^{\min\{m,n\}}$ be a vector subspace. The {\bf subcode of $\C$ supported on $V$} is $$\C(V)=\{M\in\C\mid \supp(M)\subseteq V\}.$$
\end{defn}

\section{MRD codes and optimal anticodes}

The basic invariants of a rank-metric code $\C\subseteq\mat_{n\times m}(\FF_q)$ are $n$, $m$, the dimension of $\C$ as a vector space over $\FF_q$, and its minimum distance. 

\begin{defn}\label{defn:dmin}
The \textbf{minimum distance} of a rank-metric code $0\neq\C \subseteq\mat_{n\times m}(\FF_q)$ is the integer $$d_{\min}(\C)=\min\{\rk(M) \mid M \in \C, \ M \neq 0\}.$$
We define the minimum distance of the trivial code $\C=0$ as $$d_{\min}(0)=\min\{m,n\}+1.$$
\end{defn}

Sometimes, one is also interested in the maximum rank of an element of $\C$.

\begin{defn}\label{defn:maxrk}
Let $\C\subseteq\mat_{n\times m}(\FF_q)$ be a rank-metric code. The {\bf maximum rank} of $\C$ is 
$$\max\rk(\C)=\max\{\rk(M)\mid M\in\C\}.$$
\end{defn}

Analogous definitions can be given for vector rank-metric codes, using the corresponding rank distance.

\begin{defn}\label{defn:dmin_vect}
The \textbf{minimum distance} of a vector rank-metric code $0\neq C\subseteq\FF_{q^m}^n$ is 
$$d_{\min}(C)=\min\{\rk(v)  \mid v \in C, \ v\neq 0\}.$$
The minimum distance of the trivial code $C=0$ is $$d_{\min}(0)=n+1.$$
\end{defn}

It follows from Proposition~\ref{prop:lift} that $$d_{\min}(C)=d_{\min}(\Gamma(C))$$ for any vector rank-metric code $0\neq C\subseteq\FF_{q^m}^n$ and any basis $\Gamma$ of $\FF_{q^m}$ over $\FF_q$.

\begin{defn}\label{defn:maxrk_vect}
Let $C\subseteq\FF_{q^m}^n$ be a vector rank-metric code. The {\bf maximum rank} of $C$ is 
$$\max\rk(C)=\max\{\rk(v)\mid v\in C\}.$$
\end{defn}

Both the minimum distance and the maximum rank of a code are related to the other invariants of the code. The first inequality in the next theorem goes under the name of Singleton Bound and was proved by Delsarte 
in~\cite[Theorem~5.4]{Delsarte1978}. The second goes under the name of Anticode Bound. The Anticode Bound was proved by Meshulam in~\cite[Theorem~1]{Meshulam1985} in the square case, but one can check that the proof also works in the rectangular case. The proof by Meshulam relies heavily on a result by K{\"o}nig~\cite{Koenig1931} (see also~\cite[Theorem~5.1.4]{Hall1967}). Finally, a coding-theoretic proof of the Anticode Bound was given by Ravagnani in~\cite[Proposition~47]{Ravagnani2016}.

\begin{thm}\label{thm:singleton&anticode}
Let $\C\subseteq\mat_{n\times m}(\FF_q)$ be a rank-metric code. Then
$$\dim(\C)\leq\max\{n,m\}(\min\{m,n\}-d_{\min}(\C)+1)$$ and 
$$\dim(\C)\leq\max\{n,m\}\cdot\max\rk(\C).$$
\end{thm}

\begin{rmk}
For a vector rank-metric code $C\subseteq\FF_{q^m}^n$, the Singleton Bound can be stated as 
$$\dim_{\FF_{q^m}}(C)\leq n-d_{\min}(C)+1.$$ This bound appeared in~\cite[Corollary of Lemma~1]{Gabidulin1985}, under the assumption that $n\leq m$. However, it is easy to check that the bound holds for any $n,m$.
\end{rmk}

\begin{rmk}\label{rmk:antibd}
For a vector rank-metric code $C\subseteq\FF_{q^m}^n$ with $\dim_{\FF_{q^m}}(C)\leq m$, the Anticode Bound can be stated as 
\begin{equation}\label{eqn:antbd}
\dim_{\FF_{q^m}}(C)\leq \max\rk(C).
\end{equation} 
The bound was proved in~\cite[Proposition~11]{Ravagnani2016u}, under the assumption that $n\leq m$. Using the same type of arguments however, one can easily prove the bound in the more general form stated here. Notice moreover that, if $\dim_{\FF_{q^m}}(C)>m$, then $$\dim_{\FF_{q^m}}(C)>m\geq\max\rk(C).$$ In particular, the inequality (\ref{eqn:antbd}) cannot hold if $\dim_{\FF_{q^m}}(C)>m$.
\end{rmk}

The codes whose invariants meet the bounds of Theorem~\ref{thm:singleton&anticode} go under the names of MRD codes and optimal anticodes, respectively. They have both been extensively studied. 

\begin{defn}
A vector rank-metric code $C\subseteq\FF_{q^m}^n$ is a {\bf Maximum Rank Distance (MRD) code} if 
$$\dim_{\FF_{q^m}}(C)=n-d_{\min}(C)+1.$$ 
It is an {\bf optimal vector anticode} if $$\dim_{\FF_{q^m}}(C)=\max\rk(C).$$
A rank-metric code $\C\subseteq\mat_{n\times m}(\FF_q)$ is a {\bf Maximum Rank Distance (MRD) code} if 
$$\dim(\C)=\max\{n,m\}(\min\{m,n\}-d_{\min}(\C)+1).$$ 
It is an {\bf optimal anticode} if $$\dim(\C)=\max\{n,m\}\cdot\max\rk(\C).$$
\end{defn}

\begin{rmk}\label{rmk:antitransp}
Let $\C\subseteq\mat_{n\times m}(\FF_q)$ be a rank-metric code and let $\C^\textsf{T}\subseteq\mat_{m\times n}(\FF_q)$ be the code obtained from $\C$ by transposition.
The following are immediate consequences of the definitions:
\begin{itemize}
\item $\C$ is MRD if and only if $\C^\textsf{T}$ is MRD,
\item $\C$ is an optimal anticode if and only if $\C^\textsf{T}$ is an optimal anticode.
\end{itemize}
\end{rmk}

Several examples and constructions of MRD codes are given in Chapter~\ref{LeoStorme'sChapter}. 

\begin{rmk}\label{rmk:noMRD}
Notice that $0$ and $\FF_{q^m}^n$ are the only MRD vector rank-metric codes, if $n>m$.
In fact, if a vector rank-metric code $C$ exists with 
$$\dim_{\FF_{q^m}}(C)=n-d_{\min}(C)+1<n,$$ then 
$d_{\min}(C)>1$. Assume that $C\neq 0$ and let $\Gamma$ be a basis of $\FF_{q^m}$ over $\FF_q$. 
Recall that $\dim(\Gamma(C))=m\dim_{\FF_{q^m}}(C)$ and $d_{\min}(C)=d_{\min}(\Gamma(C))$.
Then $\Gamma(C)$ is a rank-metric code of dimension 
$$\dim(\Gamma(C))=m(n-d_{\min}(\Gamma(C))+1)\leq n(m-d_{\min}(\Gamma(C))+1),$$ contradicting the assumption that $n>m$.
\end{rmk}

It is easy to produce examples of optimal anticodes and optimal vector anticodes.

\begin{ex}[Standard optimal anticodes]\label{ex:anti}
If $n\leq m$, let $\C\subseteq\mat_{n\times m}(\FF_q)$ consist of the matrices whose last $n-k$ rows are zero. Then $\dim(\C)=mk$ and $\max\rk(\C)=k$; hence $\C$ is an optimal anticode.

If $n\geq m$, let $\C\subseteq\mat_{n\times m}(\FF_q)$ consist of the matrices whose last $m-k$ columns are zero. Then $\dim(\C)=nk$ and $\max\rk(\C)=k$; hence $\C$ is an optimal anticode.
\end{ex}

\begin{ex}[Standard optimal vector anticodes]\label{ex:vanti}
Let $0\leq k\leq m$ and let $C=\langle e_1,\ldots,e_k\rangle\subseteq\FF_{q^m}^n$, where $e_i$ denotes the $i$-th vector of the standard basis of $\FF_{q^m}^n$. Then $\dim(C)=k$ and $\max\rk(C)=k$; hence $C$ is an optimal vector anticode.
\end{ex}

\begin{rmk}\label{rmk:ova}
Notice that every optimal vector anticode $C\subseteq\FF_{q^m}^n$ has 
$$\dim_{\FF_{q^m}}(C)=\max\rk(C)\leq \min\{m,n\}.$$
In particular, $\FF_{q^m}^n$ is not an optimal vector anticode if $n>m$.
\end{rmk}

It is natural to ask whether the rank-metric code associated to an MRD vector rank-metric code or to an optimal vector anticode is an MRD rank-metric code or an optimal anticode, respectively. It is easy to show that, up to the trivial exceptions, this happens only if $n\leq m$.

\begin{prop}
Let $C\subseteq\FF_{q^m}^n$ be a vector rank-metric code with $d_{\min}(C)=d$. Let $\Gamma$ be a basis of $\FF_{q^m}$ over $\FF_q$ and let $\Gamma(C)\subseteq\mat_{n\times m}(\FF_q)$ be the rank-metric code associated to $C$ with respect to $\Gamma$. If $n\leq m$, then:
\begin{enumerate}
\item $C$ is MRD if and only if $\Gamma(C)$ is MRD.
\item $C$ is an optimal vector anticode if and only if $\Gamma(C)$ is an optimal anticode.
\end{enumerate}
If $n>m$, then:
\begin{enumerate}
\item The codes $C$ and $\Gamma(C)$ are both MRD if and only if $C=0$ or $C=\FF_{q^m}^n$.
\item $C$ is an optimal vector anticode and $\Gamma(C)$ is an optimal anticode if and only if $C=0$.
\end{enumerate}
\end{prop}

\noindent\textbf{Proof:}
Notice that if $C=0$ or $C=\FF_{q^m}^n$, then both $C$ and $\Gamma(C)$ are MRD for any $n,m$. Moreover, $C=0$ is an optimal vector anticode and $\Gamma(C)=0$ is an optimal anticode for any $n,m$. In the sequel, we suppose that $C\neq 0$. Recall that for any $m,n$ one has $\dim(\Gamma(C))=m\dim_{\FF_{q^m}}(C)$, $d_{\min}(\Gamma(C))=d$, and $\max\rk(\Gamma(C))=\max\rk(C)$.

Suppose that $n\leq m$. Then:

(1) The code $C$ is MRD if and only if $\dim_{\FF_q^m}(C)=n-d+1$. 
The code $\Gamma(C)$ is MRD if and only if $\dim(\Gamma(C))=m(n-d+1)$. 
Then $C$ is MRD if and only if $\Gamma(C)$ is MRD.

(2) The code $C$ is an optimal vector anticode if and only if $\dim_{\FF_q^m}(C)=\max\rk(C)$. 
The code $\Gamma(C)$ is an optimal anticode if and only if $\dim(\Gamma(C))=m\max\rk(C)$. 
Then $C$ is an optimal vector anticode if and only if $\Gamma(C)$ is an optimal anticode.

Suppose now that $n>m$. Then:

(1) $C=0$ and $C=\FF_{q^m}^n$ are the only MRD vector rank-metric codes by Remark~\ref{rmk:noMRD}. 
The associated code $\Gamma(C)$ is MRD in both cases.

(2) The code $C$ is an optimal vector anticode if and only if $\dim_{\FF_q^m}(C)=\max\rk(C)$. 
The code $\Gamma(C)$ is an optimal anticode if and only if $\dim(\Gamma(C))=n\max\rk(C)$. 
Then $C$ is an optimal vector anticode and $\Gamma(C)$ is an optimal anticode if and only if $n\max\rk(C)=m\max\rk(C)$. Since $n>m$, this is equivalent to $\max\rk(C)=0$.\hfill$\Box$

Optimal anticodes were characterized by de Seguins Pazzis, who proved that, up to code equivalence, they are exactly the standard optimal anticodes of Example~\ref{ex:anti}.

\begin{thm}[\cite{deSeguinsPazzis2015}, Theorem 4 and Theorem 6]\label{thm:optanti}
The optimal anticodes of $\mat_{n\times m}(\FF_q)$ with respect to the rank metric are exactly the following codes:
\begin{itemize}
\item $\mat_{n\times m}(\FF_q)(V)$ for some $V\subseteq\FF_q^{\min\{m,n\}}$, if $m\neq n$,
\item $\mat_{n\times n}(\FF_q)(V)$ and $\mat_{n\times n}(\FF_q)(V)^\textsf{T}$  for some $V\subseteq\FF_q^{n}$, if $m=n$.
\end{itemize}
In particular, every optimal anticode is equivalent to a standard optimal anticode.
\end{thm}

\noindent\textbf{Proof:}
The only part of the statement which is not contained in the proof of~\cite[Theorem 4 and Theorem 6]{deSeguinsPazzis2015} is the claim that every optimal anticode is equivalent to a standard optimal anticode.
Notice that the standard optimal anticodes are $\mat_{n\times m}(\FF_q)(E_\ell)$ and $\mat_{n\times n}(\FF_q)(E_\ell)^\textsf{T}$, where $E_\ell=\langle e_1,\ldots,e_\ell\rangle$ and $\ell=0,\ldots,\min\{m,n\}$. 

If $n\leq m$ let $A\in\gln(\FF_q)$ be a matrix whose first $k=\dim(V)$ columns are a basis of $V$ and let $B\in\glm(\FF_q)$ be any matrix. If $n>m$, let $A\in\gln(\FF_q)$ be any matrix and let $B\in\glm(\FF_q)$ be a matrix whose first $k=\dim(V)$ rows are a basis of $V$.  
Then \begin{equation}\label{eqn:equiv}
A\mat_{n\times m}(\FF_q)(E_k)B=\mat_{n\times m}(\FF_q)(V).\end{equation} 
In fact, if $n\leq m$ and $\colsp(M)\subseteq E_k$, then $$\colsp(AMB)=\colsp(AM)\subseteq V.$$ Similarly, if $n>m$ and $\rowsp(M)\subseteq E_k$, then $$\rowsp(AMB)=\rowsp(MB)\subseteq V.$$ Therefore, $AMB\in\mat_{n\times m}(\FF_q)(V)$ for every $M\in\mat_{n\times m}(\FF_q)(E_k)$. Since the two vector spaces in (\ref{eqn:equiv}) have the same dimension and one is a subset of the other, they must be equal. Hence $\mat_{n\times m}(\FF_q)(V)$ is equivalent to the standard optimal anticode $\mat_{n\times m}(\FF_q)(E_k)$.
Moreover, if $n=m$, by taking the transpose of (\ref{eqn:equiv}) one has $$B^\textsf{T}\mat_{n\times n}(\FF_q)(E_k)^\textsf{T}A^\textsf{T}=\mat_{n\times n}(\FF_q)(V)^\textsf{T}.$$ Therefore $\mat_{n\times n}(\FF_q)(V)^\textsf{T}$ is equivalent to the standard optimal anticode $\mat_{n\times n}(\FF_q)(E_k)^\textsf{T}$.\hfill$\Box$

Optimal vector anticodes were characterized by Ravagnani in~\cite[Theorem~18]{Ravagnani2016u}, under the assumption that $n\leq m$. One can also show that, up to code equivalence, optimal vector anticodes are exactly the standard optimal vector anticodes of Example~\ref{ex:vanti}. 
Notice that a vector rank-metric code $C\subseteq\FF_{q^m}^n$ with $\dim_{\FF_{q^m}}(C)>m$ cannot be an optimal vector anticode by Remark~\ref{rmk:ova}. Hence we may assume without loss of generality that $\dim_{\FF_{q^m}}(C)\leq m$. 

\begin{thm}
Let $C\subseteq\FF_{q^m}^n$ be a vector rank-metric code with $k=\dim_{\FF_{q^m}}(C)\leq m$. 
The following are equivalent:
\begin{enumerate}
\item $C$ is an optimal vector anticode,
\item $C$ has a basis consisting of vectors with entries in $\FF_q$,
\item $C\sim\langle e_1,\ldots,e_k\rangle$, where $e_i$ denotes the $i$-th vector of the standard basis of $\FF_{q^m}^n$.
\end{enumerate}
\end{thm}

\noindent\textbf{Proof:}
Let $\phi:\FF_{q^m}^n\to\FF_{q^m}^n$ be the Frobenius endomorphism, defined by $\phi(v_1,\ldots,v_n)=(v_1^q,\ldots,v_n^q)$. Recall that a subspace $V\subseteq\FF_{q^m}^n$ is fixed by $\phi$ if and only if it has an $\FF_{q^m}$-basis that consists of vectors with entries in $\FF_q$. Combining this fact with the argument in~\cite[Theorem~18]{Ravagnani2016u}, one has that $C$ is an optimal vector anticode if and only if $C$ has a basis consisting of vectors with entries in $\FF_q$. Although~\cite[Theorem~18]{Ravagnani2016u} is proved under the assumption that $n\leq m$, one can check that the proof works for arbitrary $n,m$, under the assumption that $\dim_{\FF_{q^m}}(C)\leq m$. This proves that (1) and (2) are equivalent. 

By Theorem~\ref{thm:berger}, $C\sim\langle e_1,\ldots,e_k\rangle$ if and only if there exist $\alpha\in\FF_{q^m}^*$ and $B\in\glm(\FF_q)$ such that $$C=\alpha\langle e_1,\ldots,e_k\rangle B=\langle e_1,\ldots,e_k\rangle B.$$  Equivalence of (2) and (3) follows readily.\hfill$\Box$

\section{Duality and MacWilliams Identities}\label{sect:MacWill}

The usual scalar product for matrices induces a notion of dual for rank-metric codes.

\begin{defn}\label{defn:dual}
The \textbf{dual} of $\C\subseteq\mat_{n\times m}(\FF_q)$ is 
$$\mathcal{C}^\perp=\{M \in\mat_{n\times m}(\FF_q) \mid \tr(MN^\textsf{T})=0 \text{ for all }N\in\mathcal{C}\},$$ where $\tr(\cdot)$ denotes the trace of a matrix. 
\end{defn}

The usual scalar product of $\FF_{q^m}^n$ induces a notion of dual for vector rank-metric codes.

\begin{defn}
The \textbf{dual} of $C\subseteq\FF_{q^m}^n$ is the vector rank-metric code $$C^{\perp}:= \{v \in \FF_{q^m}^n \mid \langle v, w\rangle =0 \mbox{ for all } w \in C \},$$ where $\langle \cdot , \cdot \rangle$ is the standard inner product of $\FF_{q^m}^n$. 
\end{defn}

The two notions of dual code are compatible with the definition of associated rank-metric code with respect to a basis of $\FF_{q^m}$ over $\FF_q$, for a suitable choice of bases.

\begin{defn}\label{defn:ortbases}
Two bases $\Gamma=\{\gamma_1,\ldots,\gamma_m\}$ and $\Gamma'=\{\gamma'_1,\ldots,\gamma'_m\}$ of $\FF_{q^m}$ over $\FF_q$ are {\bf orthogonal} if $$\tr(\gamma_i\gamma'_j)=\left\{\begin{array}{ll}
1 & \mbox{if }i=j\\
0 & \mbox{if } i\neq j
\end{array}\right.$$
where $\tr(\cdot)$ denotes the trace relative to the field extension $\FF_{q^m}\supseteq\FF_q$.
\end{defn}

\begin{prop}[\cite{Ravagnani2016}, Theorem~21]\label{prop:dual}
Let $C\subseteq\FF_{q^m}^n$ be a vector rank-metric code and let $\Gamma,\Gamma'$ be orthogonal bases of $\FF_{q^m}$ over $\FF_q$. Then $$\Gamma(C)^\perp=\Gamma'(C^\perp).$$
\end{prop}

We now give a simple example to illustrate Proposition~\ref{prop:dual}.

\begin{ex}
Let $C$ be the vector rank-metric code $C=\langle(1,\alpha)\rangle\subseteq\FF_8^2$, where $\FF_8=\FF_2[\alpha]/(\alpha^3+\alpha+1)$. Its dual code is $C^\perp=\langle(1,\alpha^2+1)\rangle\subseteq\FF_8^2.$\vskip .02cm\noindent
Let $\Gamma=\{\gamma_1=1,\gamma_2=\alpha,\gamma_3=\alpha^2\}$ be an $\FF_2$-basis of $\FF_8$. The rank-metric code associated to $C$ with respect to $\Gamma$ is 
$$\Gamma(C)=\left\langle\begin{pmatrix} 1 & 0 & 0 \\ 0 & 1 & 0 \end{pmatrix}, 
\begin{pmatrix} 0 & 1 & 0 \\ 0 & 0 & 1 \end{pmatrix}, \begin{pmatrix} 0 & 0 & 1 \\ 1 & 1 & 0  \end{pmatrix}
\right\rangle\subseteq\mat_{2\times 3}(\FF_2).$$ Its dual code is 
$$\Gamma(C)^\perp=\left\langle\begin{pmatrix} 0 & 0 & 1 \\ 1 & 0 & 0 \end{pmatrix}, 
\begin{pmatrix} 1 & 0 & 1 \\ 0 & 1 & 0 \end{pmatrix}, \begin{pmatrix} 0 & 1 & 0 \\ 0 & 0 & 1 \end{pmatrix}
\right\rangle\subseteq\mat_{2\times 3}(\FF_2).$$
The orthogonal basis of $\Gamma$ is $\Gamma'=\{\gamma_1'=1,\gamma_2'=\alpha^2,\gamma_3'=\alpha\}$.
The rank-metric code associated to $C^\perp$ with respect to $\Gamma'$ is $$\Gamma'(C^\perp)=\left\langle\begin{pmatrix} 1 & 0 & 0 \\ 1 & 1 & 0 \end{pmatrix}, \begin{pmatrix} 0 & 0 & 1 \\ 1 & 0 & 0 \end{pmatrix},
\begin{pmatrix} 0 & 1 & 0 \\ 0 & 0 & 1 \end{pmatrix}\right\rangle.$$
It is easy to check that $\Gamma(C)^\perp=\Gamma'(C^\perp)$.
\end{ex}

There are a number of interesting relations between the invariants of a code and those of its dual. The simplest one is probably the equality $$\dim(\C)+\dim(\C^\perp)=mn,$$ which holds for any rank-metric code $\C\subseteq\mat_{n\times m}(\FF_q)$. 

The minimum distances of $\C$ and $\C^\perp$ do not satisfy such a simple relation. Nevertheless, one can relate them through the next inequality, which follows easily from the Singleton Bound.

\begin{prop}[\cite{Ravagnani2016}, Proposition~43]\label{prop:dualdistance}
Let $\C\subseteq\mat_{n\times m}(\FF_q)$ be a rank-metric code. Then 
$$d_{\min}(\C)+d_{\min}(\C^{\perp})\leq\min\{m,n\}+2$$ and equality holds if and only if $\C$ is MRD.
\end{prop}

Using the Anticode Bound, one can produce an inequality which involves $\max\rk(\C)$ and $\max\rk(\C^\perp)$.

\begin{prop}[\cite{Ravagnani2016}, Proposition~55]
Let $\C\subseteq\mat_{n\times m}(\FF_q)$ be a rank-metric code. Then 
$$\max\rk(\C)+\max\rk(\C^\perp)\geq\min\{m,n\}$$ and equality holds if and only if $\C$ is an optimal anticode.
\end{prop}

Finally, by combining the Singleton Bound and the Anticode Bound, one obtains an inequality which involves $d_{\min}(C)$ and $\max\rk(\C^\perp)$.

\begin{prop}[\cite{Ravagnani2016}, Proposition~49]\label{prop:MRDanti}
Let $\C\subseteq\mat_{n\times m}(\FF_q)$ be a rank-metric code. Then 
$$d_{\min}(\C)\leq\max\rk(\C^\perp)+1.$$
\end{prop}

Notice that equality holds in Proposition~\ref{prop:MRDanti} if and only if $\C$ is MRD and an optimal anticode. We will see in Corollary~\ref{cor:MRD&optanti} that this is the case if and only if $\C=0$ or $\C=\mat_{n\times m}(\FF_q)$.

Although the minimum distance of $\C$ and $\C^\perp$ do not determine each other, the weight distribution of $\C$ determines the weight distribution of $\C^\perp$, and vice versa. We now define the weight distribution, which is an important invariant of a rank-metric code.

\begin{defn}\label{defn:weightdistr}
Let $\C\subseteq\mat_{n\times m}(\FF_q)$ be a rank-metric code. The {\bf weight distribution} of $\C$ is the collection of natural numbers $$A_i(\C)=\#\{M\in\C\mid \rk(M)=i\},\; i=0,\ldots,\min\{m,n\}.$$
\end{defn}

Clearly $A_0(\C)=1$, $d_{\min}(\C)=\min\{i\mid A_i(\C)\neq 0, i\neq 0\}$, and $\max\rk(\C)=\max\{i\mid A_i(\C)\neq 0\}$. 

\begin{defn}
The {\bf $q$-ary Gaussian coefficient} of $a,b\in\ZZ$ is $$\begin{bmatrix} a \\ b\end{bmatrix}_q=
\left\{\begin{array}{cl} 
0 & \mbox{ if $a < 0, \; b < 0$, or $b >a$,} \\ 
1 & \mbox{ if $b=0$ and $a \geq 0$,} \\
\frac{(q^a-1)(q^{a-1}-1) \cdots (q^{a-b+1}-1)}{(q^b-1)(q^{b-1}-1) \cdots (q-1)} & \mbox{ otherwise.} 
\end{array}\right.$$
\end{defn}

The relations between the weight distribution of $\C$ and $\C^\perp$ go under the name of MacWilliams Identities, and were first proved in~\cite[Theorem~3.3]{Delsarte1978}. A different proof, inspired by~\cite[Theorem~27]{Ravagnani2018}, was given in~\cite[Theorem~2]{Gorla2018ch}. Another proof was given in~\cite[Proposition~15]{Shiromoto2018}. 

\begin{thm}[MacWilliams Identities]
Let $\C\subseteq\mat_{n\times m}(\FF_q)$ be a rank-metric code. One has 
$$A_i(\C^\perp)=\frac{1}{|\C|}\sum_{j=0}^{\min\{m,n\}} A_j(\C)\sum_{\ell=0}^{\min\{m,n\}} (-1)^{i-\ell} q^{\max\{m,n\} k+{i-\ell\choose 2}}\begin{bmatrix}
\min\{m,n\}-\ell \\ \min\{m,n\}-i\end{bmatrix}_q \begin{bmatrix} \min\{m,n\}-j \\ \ell\end{bmatrix}_q$$ for $i=0,\ldots,\min\{m,n\}$.
\end{thm}

The following is an equivalent formulation of the MacWilliams Identities. Identities of this form for vector rank-metric codes were proved in~\cite[Proposition~3]{Gadouleau2007}. The same identities were proved in~\cite[Theorem~21]{Ravagnani2016} for rank-metric codes. 

\begin{thm}
Let $\C\subseteq\mat_{n\times m}(\FF_q)$ be a rank-metric code. One has 
$$\sum_{i=0}^{\min\{m,n\}-\ell} A_i(\C)\begin{bmatrix} \min\{m,n\}-i \\ \ell \end{bmatrix}_q=
\frac{|\C|}{q^{\max\{m,n\}\cdot\ell}}\sum_{j=0}^\ell A_i(\C^\perp)\begin{bmatrix} \min\{m,n\}-j \\ \ell-j \end{bmatrix}_q$$
for $\ell=0,\ldots,\min\{m,n\}$.
\end{thm}

From the MacWilliams Identities, one can derive a number of nontrivial consequences. 
Here we give two relevant ones, starting with the celebrated result of Delsarte, which states that the dual of an MRD code is MRD.

\begin{thm}[\cite{Delsarte1978}, Theorem~5.5]\label{thm:dualMRD}
Let $\C\subseteq\mat_{n\times m}(\FF_q)$ be a rank-metric code. Then $\C$ is MRD if and only if $\C^\perp$ is MRD.
\end{thm}

The weight distribution of an MRD code was first computed by Delsarte in~\cite[Theorem~5.6]{Delsarte1978} and can be derived from the MacWilliams Identities via a standard computation. An analogous result can be obtained for dually quasi-MRD codes. 

\begin{defn}\index{code!dually quasi-MRD}
Let $\C \subseteq \mat_{n\times m}(\FF_q)$ be a rank-metric code. $\C$ is {\bf dually quasi-MRD} if 
$$d_{\min}(\C)+d_{\min}(\C^{\perp})=\min\{m,n\}+1.$$
\end{defn}

Discussing the family of dually quasi-MRD codes is beyond the scope of this chapter. The definition however is motivated by Proposition~\ref{prop:dualdistance}, which shows that dually quasi-MRD codes are exactly the non-MRD codes which maximize the quantity $d_{\min}(\C)+d_{\min}(\C^{\perp})$. We refer the interested reader to~\cite{DeLaCruz2018} for a discussion of the properties of codes which are close to being MRD in this sense. 
The weight distribution of a dually quasi-MRD code was computed in~\cite[Corollary~28]{DeLaCruz2018}. Below we give a statement that covers both MRD and  dually quasi-MRD codes.

\begin{thm}\label{thm:rkMRD}
Let $\C\subseteq\mat_{n\times m}(\FF_q)$ be an MRD or dually quasi-MRD rank-metric code. Let $d=d_{\min}(\C)$. Then $A_0(\C)=1$, $A_i(\C)=0$ for $i=1,\ldots,d-1$, and
$$A_i(\C)=\begin{bmatrix} \min\{m,n\} \\ i \end{bmatrix}_q \sum_{j=0}^{i-d}(-1)^{j}q^{{j\choose 2}}
\begin{bmatrix} i \\ j \end{bmatrix}_q \left(q^{\dim(\C)-\max\{m,n\}(\min\{m,n\}+j-i)}-1\right)$$
for $i=d,\ldots,\min\{m,n\}.$
\end{thm}

The analogue of Theorem~\ref{thm:dualMRD} for optimal anticodes was proved by Ravagnani.

\begin{thm}[\cite{Ravagnani2016}, Theorem~54]
Let $\C\subseteq\mat_{n\times m}(\FF_q)$ be a rank-metric code. Then $\C$ is an optimal anticode if and only if $\C^\perp$ is an optimal anticode.
\end{thm}

Ravagnani also proved the next interesting result, relating MRD codes and optimal anticodes.

\begin{prop}[\cite{Ravagnani2016}, Proposition~53]
Let $\C\subseteq\mat_{n\times m}(\FF_q)$ be a rank-metric code of dimension $\dim(\C)=k\cdot \max\{m,n\}$. Then $\C$ is an optimal anticode if and only if $$\C+\D=\mat_{n\times m}(\FF_q)$$ for every $\D\subseteq\mat_{n\times m}(\FF_q)$ MRD code of minimum distance $d_{\min}(\D)=k+1$.
\end{prop}

\section{Generalized weights}\label{sect:genwgts}

Generalized Hamming weights were introduced by Helleseth, Kl{\o}ve, and Mykkeltveit in~\cite{Helleseth1977} for  linear block codes. In~\cite{Wei1991}, Wei studied them in the context of wire-tap channels.
Different definitions of generalized weights were given for vector rank-metric codes and rank-metric codes. In this section, we give the different definitions and compare them with each other.

In the context of vector rank-metric codes, generalized weights were first defined by Oggier and Sboui. 

\begin{defn}[\cite{Oggier2012}, Definition~1]\label{defn:higherwts_vect1}
Let $n\leq m$ and let $C\subseteq\FF_{q^m}^n$ be a vector rank-metric code. 
The {\bf generalized weights} of $C$ are 
$$\begin{aligned} w_i(C)=\min_D\{\max_v\{\dim\supp(v)\mid v\in D, v\neq 0\}\mid D\subseteq C, \dim_{\FF_{q^m}}(D)=i\},\;\\
i=1,\ldots,\dim_{\FF_{q^m}}(C).\end{aligned}$$
\end{defn}

A definition of relative generalized weights for vector rank-metric codes was given by Kurihara, Matsumoto, and Uyematsu. Let $\phi:\FF_{q^m}^n\to\FF_{q^m}^n$ be the {\bf Frobenius endomorphism}, 
defined by $\phi(v_1,\ldots,v_n)=(v_1^q,\ldots,v_n^q)$. 

\begin{defn}[\cite{Kurihara2015}, Definition~2]\label{defn:relhigherwts_vect}
Let $C\subseteq\FF_{q^m}^n$ be a vector rank-metric code, and let $D\subsetneq C$ be a proper subcode.
Let $\phi:\FF_{q^m}^n\to\FF_{q^m}^n$ be the Frobenius endomorphism.
The {\bf relative generalized weights} of $C$ and $D$ are 
$$\begin{aligned}
w_i(C,D)=\min\{\dim\supp(V)\mid V\subseteq\FF_{q^m}^n,\; \phi(V)=V,\; \dim_{\FF_{q^m}}(C\cap V)-\dim_{\FF_{q^m}}(D\cap V)\geq i\},\\ i=1,\ldots,\dim_{\FF_{q^m}}(C)-\dim_{\FF_{q^m}}(D).\end{aligned}$$
\end{defn}

The relative generalized weights of $C$ and $0$ are by definition 
$$\begin{aligned} w_i(C,0)=\min\{\dim\supp(V)\mid V\subseteq\FF_{q^m}^n,\; \phi(V)=V,\; 
\dim_{\FF_{q^m}}(C\cap V)\geq i\},\;\; \\ i=1,\ldots,\dim_{\FF_{q^m}}(C).\end{aligned}$$

In~\cite{Ducoat2015}, Ducoat proposed and studied the following modification of Definition~\ref{defn:higherwts_vect1}. For any $D\subseteq\FF_{q^m}^n$, let $D^*=D+\phi(D)+\cdots+\phi^{m-1}(D)$, where $\phi$ denotes the Frobenius endomorphism. $D^*$ is the smallest $\FF_{q^m}$-linear space containing $D$ which is fixed by $\phi$.

\begin{defn}\label{defn:higherwts_vect2}
Let $C\subseteq\FF_{q^m}^n$ be a vector rank-metric code. 
The {\bf generalized weights} of $C$ are 
$$\begin{aligned}w_i(C)=\min_D\{\max_v\{\dim\supp(v)\mid v\in D^*, v\neq 0\}\mid D\subseteq C, \dim_{\FF_{q^m}}(D)=i\},\;\\
i=1,\ldots,\dim_{\FF_{q^m}}(C).\end{aligned}$$
\end{defn}

Notice that, although the definition by Ducoat does not assume $n\leq m$, most of the results that he establishes in~\cite{Ducoat2015} do.

It was shown by Ducoat in~\cite[Proposition~II.1]{Ducoat2015} for $n\leq m$, and by Jurrius and Pellikaan in~\cite[Theorem~5.4]{Jurrius2017} for any $n,m$, that the relative generalized weights of $C$ and $0$ agree with the generalized weights of $C$, i.e., 
$$w_i(C,0)=w_i(C),\; \mbox{ for } i=1,\ldots,\dim_{\FF_{q^m}}(C).$$ Moreover, it follows from~\cite[Theorem~5.2 and Theorem~5.8]{Jurrius2017} that Definition~\ref{defn:higherwts_vect1} and Definition~\ref{defn:higherwts_vect2} are equivalent for $n\leq m$.

The next definition is the natural analogue of the definition of generalized rank weights for linear block codes, endowed with the Hamming distance. It was given by Jurrius and Pellikaan, who in~\cite[Theorem~5.2]{Jurrius2017} proved that it is equivalent to Definition~\ref{defn:higherwts_vect1} if $n\leq m$. In~\cite[Theorem~5.8]{Jurrius2017} they proved that it is equivalent to Definition~\ref{defn:higherwts_vect2}. 

\begin{defn}[\cite{Jurrius2017}, Definition~2.5]\label{defn:higherwts_vect3}
Let $C\subseteq\FF_{q^m}^n$ be a vector rank-metric code.
The {\bf generalized weights} of $C$ are 
$$w_i(C)=\min\{\dim\supp(D)\mid D\subseteq C, \dim_{\FF_{q^m}}(D)=i\},\; 
i=1,\ldots,\dim_{\FF_{q^m}}(C).$$
\end{defn}

The next result provides another equivalent definition of generalized weights for vector rank-metric codes. It was proved by Ravagnani under the assumption $n\leq m$, but it can easily be extended to arbitrary $n,m$ as follows.

\begin{thm}[\cite{Ravagnani2016u}, Corollary~19]\label{thm:gwova}
Let $C\subseteq\FF_{q^m}^n$ be a vector rank-metric code of dimension $\dim_{\FF_{q^m}}(C)\leq m$. Then 
$$w_i(C)=\min\{\dim_{\FF_{q^m}}(A)\mid A\subseteq\FF_{q^m}^n \mbox{ optimal vector anticode},\; \dim_{\FF_{q^m}}(C\cap A)\geq i\}$$ 
for $i=1,\ldots,\dim_{\FF_{q^m}}(C).$
\end{thm}

\begin{rmk}
Let $C\subseteq\FF_{q^m}^n$ be a vector rank-metric code of dimension $\dim_{\FF_{q^m}}(C)>m$ 
and let $m<i\leq\dim_{\FF_{q^m}}(C)$. Then the quantity 
$$\min\{\dim_{\FF_{q^m}}(A)\mid A\subseteq\FF_{q^m}^n \mbox{ optimal vector anticode},\; 
\dim_{\FF_{q^m}}(C\cap A)\geq i\}$$ cannot be equal to $w_i(C)$, since by Remark~\ref{rmk:ova} there exists no optimal vector anticode $A$ with $\dim_{\FF_{q^m}}(A)\geq \dim_{\FF_{q^m}}(C\cap A)\geq i>m$ .
\end{rmk}

The first definition of generalized weights for the larger class of rank-metric codes was given by Ravagnani.

\begin{defn}[\cite{Ravagnani2016u}, Definition~23]\label{defn:higherwts}
Let $\C\subseteq \mat_{n\times m}(\FF_q)$ be a rank-metric code. The {\bf generalized weights} of $\C$ are
$$\begin{aligned}
d_i(\C)= \frac{1}{\max\{m,n\}} \min\{\dim(\cA) \mid \cA\subseteq \mat_{n\times m}(\FF_q) \mbox{ optimal anticode,}\; \dim(\C \cap \cA) \ge i\},\\
i=1,\ldots,\dim(\C).
\end{aligned}$$
\end{defn}

\begin{rmk}\label{rmk:higherwts&supp}
The characterization of optimal anticodes from Theorem~\ref{thm:optanti}, together with the observation that 
$$\dim(\mat_{n\times m}(\FF_q)(V))=\max\{n,m\}\cdot\dim(V),$$ implies that for $i=1,\ldots,\dim(\C)$ one has
$$d_i(\C)=\min\{\dim(V)\mid V\subseteq\FF_q^{\min\{n,m\}},\ \dim(\C(V))\geq i\}\;\;\mbox{ if } m\neq n$$
and 
$$d_i(\C)=\min\{\dim(V)\mid V\subseteq\FF_q^n,\ \max\{\dim(\C(V)),\dim(\C^\textsf{T}(V))\}\geq i\}\;\;\mbox{ if } m=n.$$
Notice that, for $m=n$, this definition of generalized weights is coherent with the definition of support given in Remark~\ref{rmk:supp_square}.
\end{rmk}

Generalized weights for vector rank-metric codes and their associated rank-metric codes are related as follows.

\begin{thm}[\cite{Ravagnani2016u}, Theorem~28]\label{thm:grwvect}
Let $n\leq m$ and let $C\subseteq\FF_{q^m}^n$ be a vector rank-metric code. Let $\Gamma=\{\gamma_1,\ldots,\gamma_m\}$ be a basis of $\FF_{q^m}$ over $\FF_q$. Then 
$$w_i(C)=d_{mi-e}(\Gamma(C))$$ for $i=1,\ldots,\dim_{\FF_{q^m}}(C)$ and $e=0,\ldots,m-1$.
\end{thm}

\begin{rmk}
In particular, under the assumptions of Theorem~\ref{thm:grwvect} one has that $$d_{m(i-1)+1}(\Gamma(C))=\ldots=d_{mi-1}(\Gamma(C))=d_{mi}(\Gamma(C))$$ for $i=1,\ldots,\dim_{\FF_{q^m}}(C)$.
\end{rmk}

One can easily find an example that shows that the equality in Theorem~\ref{thm:grwvect} does not hold if $n>m$.

\begin{ex}\label{ex:equivinvar}
Let $C=\langle(1,0,0)\rangle\subseteq\FF_4^3$, where $\FF_4=\FF_2[\alpha]/(\alpha^2+\alpha+1)$. Then $$w_1(C)=d_{\min}(C)=1.$$
Let $\Gamma=\{1,\alpha\}$ be an $\FF_2$-basis of $\FF_4$. Then $$\Gamma(C)=\left\langle
\begin{pmatrix}1 & 0 \\ 0 & 0 \\ 0 & 0\end{pmatrix}, \begin{pmatrix}0 & 1 \\ 0 & 0 \\ 0 & 0\end{pmatrix}\right\rangle$$ has $d_1(\Gamma(C))=d_{\min}(\Gamma(C))=1$. Since $\Gamma(C)$ is not an optimal anticode, any nonzero optimal anticode $\mathcal{A}\supsetneq\Gamma(C)$. Hence $\mathcal{A}$ must have $\max\rk(\mathcal{A})=2$ and $\dim(\mathcal{A})=6$. Therefore, $d_2(\Gamma(C))=2$.
\end{ex}

A definition of relative generalized weights for rank-metric codes was proposed by Mart{\'\i}nez-Pe\~{n}as and Matsumoto. This yields in particular a definition of generalized weights, which is different from Definition~\ref{defn:higherwts}, as we discuss below. In order to avoid confusion, we call the weights defined by Mart{\'\i}nez-Pe\~{n}as and Matsumoto generalized matrix weights. 

\begin{defn}[\cite{MartinezPenas2018}, Definition~10]\label{defn:higherwtsMP}
Let $\C \subseteq \mat_{n\times m}(\FF_q)$ be a rank-metric code, and let $\D\subsetneq \C$ be a proper subcode.
Denote by $$\mat_{n\times m}(\FF_q)_V^{\colsp}=\{M\in\mat_{n\times m}(\FF_q)\mid\colsp(M)\subseteq V\}.$$
The $i$-th \textbf{relative generalized matrix weight} of $\C$ and $\D$ is 
$$\begin{aligned}
\delta_i(\C,\D)=\min\{\dim(V)\mid \dim(\C\cap\mat_{n\times m}(\FF_q)_V^{\colsp})-\dim(\D\cap\mat_{n\times m}(\FF_q)_V^{\colsp})\geq i,\\ V\subseteq\FF_q^n\},\;\; i=1,\ldots,\dim(\C)-\dim(\D).\end{aligned}$$
The $i$-th \textbf{generalized matrix weight} of $\C$ is the $i$-th relative generalized matrix weight of $\C$ and $0$, i.e.,
$$\delta_i(\C)=\min\{\dim(V)\mid V\subseteq\FF_q^n,\; \dim(\C\cap\mat_{n\times m}(\FF_q)_V^{\colsp})\geq i\},\;\; 
i=1,\ldots,\dim(\C).$$
\end{defn}

Generalized matrix weights measure the information leakage to a wire-tapper in a linearly coded network and, more generally, in a matrix-multiplicative channel. The model discussed in~\cite{MartinezPenas2018} is not invariant with respect to transposition, since the wiretapper's observation is $AM$, where $M$ is the codeword and $A$ is the wiretap transfer matrix. Accordingly, in Definition~\ref{defn:higherwtsMP} the authors consider the column space of the matrix independently of whether the matrix has more rows or columns. 
In particular, one should not expect Definition~\ref{defn:higherwtsMP} to be equivalence-invariant, i.e. equivalent codes may not have the same generalized matrix weights. In Example~\ref{ex:invar} we show that this can in fact happen. Therefore, Definition~\ref{defn:higherwtsMP} is not equivalence-invariant. 
The next proposition shows that Definition~\ref{defn:higherwts} is equivalence-invariant.

\begin{prop}[\cite{Gorla2018p}, Proposition~2.4]\label{prop:equivgw}
Let $\C,\D\subseteq \mat_{n\times m}(\FF_q)$ be rank-metric codes. If $\C\sim\D$, then $$d_i(\C)=d_i(\D)\;\; \mbox{ for } i=1,\ldots,\dim(\C).$$
\end{prop}

The next result compares Definition~\ref{defn:higherwts} and Definition~\ref{defn:higherwtsMP}. It follows easily from Theorem~\ref{thm:optanti} and appears in the literature as~\cite[Theorem 9]{MartinezPenas2018}.
Notice that the assumption that $n\leq m$ is missing throughout~\cite[Section VIII.C]{MartinezPenas2018}. As a consequence, the statement of~\cite[Theorem 9]{MartinezPenas2018} claims that Definition~\ref{defn:higherwts} and Definition~\ref{defn:higherwtsMP} agreee for $m\neq n$; however the result is proved only for $n<m$.

\begin{thm}\label{prop:relngw}
Let $\C\subseteq\mat_{n\times m}(\FF_q)$ be a rank-metric code. Then:
\begin{itemize} 
\item If $m>n$, then $d_i(\C)=\delta_i(\C)$ for $i=1,\ldots,\dim(\C)$. 
\item If $m=n$, then $d_i(\C)\le\delta_i(\C)$ for $i=1,\ldots,\dim(\C)$.
\end{itemize}
\end{thm}

\noindent\textbf{Proof:}
The thesis follows from Remark~\ref{rmk:higherwts&supp}, after observing that for $n\leq m$ one has 
$$\C(V)=\C\cap\mat_{n\times m}(\FF_q)_V^{\colsp}.$$\hfill$\Box$

One can easily find examples that show that Definition~\ref{defn:higherwts} and Definition~\ref{defn:higherwtsMP} do not agree in the case $m=n$. 

\begin{ex}[\cite{Gorla2018p}, Example~2.10]\label{ex:invar}
Let $\C\subseteq\mat_{2\times 2}(\FF_2)$ be the code 
$$\C:=\left\{ \begin{pmatrix} a & a \\ b & b \end{pmatrix}  :  a,b \in \FF_2 \right\}.$$
Then $\C$ is an optimal anticode of dimension 2. Therefore
$d_2(\C)=1$. On the other hand, $\supp(\C)=\FF_2^2$; hence 
$\delta_2(\C)=2 \neq d_2(\C)$.

Moreover, observe that $\C \sim \C^\textsf{T}$.
In particular, $d_2(\C)=d_2(\C^\textsf{T})=1$. However,
$\delta_2(\C)=2$, while $\delta_2(\C^\textsf{T})=1$.
\end{ex}

In fact, one can also find examples that show that Definition~\ref{defn:higherwts} and Definition~\ref{defn:higherwtsMP} do not agree in the case $m<n$. This implies in particular that the part of the statement of~\cite[Theorem 9]{MartinezPenas2018} concerning the case $m<n$ is incorrect.

\begin{ex}\label{ex:gw}
Let $\C\subseteq\mat_{3\times 2}(\FF_2)$ be the code 
$$\C:=\left\{ \begin{pmatrix} a & a \\ b & b \\ c & c \end{pmatrix}  :  a,b,c \in \FF_2 \right\}.$$
Then $\C$ is an optimal anticode of dimension 3. Therefore
$d_3(\C)=1$. On the other hand, $$\sum_{M\in\C}\colsp(M)=\FF_2^3;$$ hence 
$\delta_3(\C)=3$.
\end{ex}

The code $\C\subseteq\mat_{3\times 2}(\FF_2)$ of Example~\ref{ex:gw} has $d_3(\C)<\delta_3(\C)$. One may wonder whether $d_i(\C)\leq\delta_i(\C)$ for all $i$, for a rank-metric code $\C\subseteq\mat_{n\times m}(\FF_q)$ with $n>m$. The answer turns out to be negative, as the next example shows. 

\begin{ex}
Let $\C\subseteq\mat_{3\times 2}(\FF_2)$ be the code 
$$\C:=\left\{ \begin{pmatrix} a & b \\ 0 & 0 \\ 0 & 0 \end{pmatrix}  :  a,b \in \FF_2 \right\}.$$
In Example~\ref{ex:equivinvar} we showed that $d_2(\C)=2$.
On the other hand, $$\sum_{M\in\C}\colsp(M)=\langle(1,0,0)\rangle;$$ hence 
$\delta_2(\C)=1$.
\end{ex}

In fact, as an easy consequence of Theorem~\ref{thm:optanti} and of Remark~\ref{rmk:antitransp} one obtains the following result.

\begin{thm}
Let $\C\subseteq\mat_{n\times m}(\FF_q)$ be a rank-metric code. Then:
\begin{itemize} 
\item If $m<n$, then $d_i(\C)=\delta_i(\C^\textsf{T})$ for $i=1,\ldots,\dim(\C)$. 
\item If $m=n$, then $d_i(\C)=\min\{\delta_i(\C),\delta_i(\C^\textsf{T})\}$ for $i=1,\ldots,\dim(\C)$.
\end{itemize}
\end{thm}

\noindent\textbf{Proof:}
Since $n\geq m$, then 
\begin{equation}\label{eqn:CT}
\C^\textsf{T}(V)=\C^\textsf{T}\cap\mat_{m\times n}(\FF_q)_V^{\colsp}.
\end{equation}
If $n>m$, then 
$$d_i(\C)=d_i(\C^\textsf{T})=\min\{\dim(V)\mid V\subseteq\FF_q^m,\ \dim(\C^\textsf{T}(V))\geq i\}=\delta_i(\C^\textsf{T}),$$
where the first equality follows from Proposition~\ref{prop:equivgw}, the second from Remark~\ref{rmk:higherwts&supp}, and the third from (\ref{eqn:CT}).
If $n=m$, then
$$d_i(\C)=\min\{\dim(V)\mid V\subseteq\FF_q^n,\ \max\{\dim(\C(V)),\dim(\C^\textsf{T}(V))\}\geq i\}=\min\{\delta_i(\C),\delta_i(\C^\textsf{T})\},$$
where the first equality follows from Remark~\ref{rmk:higherwts&supp} and the second from (\ref{eqn:CT}).\hfill$\Box$

As in the case of generalized weights, one can relate the generalized weights of a vector rank-metric code and the generalized matrix weights of its associated rank-metric code. In fact more is true, since the relative versions of the weights can also be related, and the assumption $n\leq m$ is not needed.
The proof of the next result is essentially the same as the proof of Theorem~\ref{thm:grwvect}.

\begin{thm}[\cite{MartinezPenas2018}, Theorem~7]
Let $C\subseteq\FF_{q^m}^n$ be a vector rank-metric code and let $D\subsetneq C$ be a proper subcode. 
Let $\Gamma=\{\gamma_1,\ldots,\gamma_m\}$ be a basis of $\FF_{q^m}$ over $\FF_q$. Then 
$$w_i(C,D)=\delta_{mi-e}(\Gamma(C),\Gamma(D))$$ for $i=1,\ldots,\dim_{\FF_{q^m}}(C)-\dim_{\FF_{q^m}}(D)$ and $e=0,\ldots,m-1$.
\end{thm}

We conclude this section with a few results on generalized weights. The next theorem establishes some properties of the sequence of generalized weights of a rank-metric code.

\begin{thm}[\cite{Ravagnani2016u}, Theorem~30]\label{thm:weightsineq}
Let $\C\subseteq \mat_{n\times m}(\FF_q)$ be a rank-metric code of dimension $\dim(\C)=\ell$. Then:
\begin{enumerate}
\item $d_1(\C)=d_{\min}(\C)$,
\item $d_\ell(\C)\leq\min\{m,n\}$,
\item $d_i(\C)\leq d_{i+1}(\C)$ for $i=1,\ldots,\ell-1$,
\item $d_i(\C)\leq d_{i+\max\{m,n\}}(\C)$ for $i=1,\ldots,\ell-\max\{m,n\}$.
\end{enumerate}
\end{thm}

Theorem~\ref{thm:weightsineq} allows one to compute the generalized weights of MRD codes and optimal anticodes. In Section~\ref{sect:MacWill} we stated analogous results for the weight distribution of MRD codes and optimal anticodes.

\begin{cor}[\cite{Ravagnani2016u}, Corollary~31]\label{cor:wtsMRD}
Let $\C\subseteq \mat_{n\times m}(\FF_q)$ be a rank-metric code of dimension $\dim(\C)=\ell=k\cdot\max\{m,n\}$. 
The following are equivalent:
\begin{itemize}
\item $\C$ is MRD,
\item $d_i(\C)=\min\{m,n\}-k+\lceil i/\max\{m,n\} \rceil$ for $i=1,\ldots,\ell$.
\end{itemize}
\end{cor}

\begin{cor}[\cite{Ravagnani2016u}, Corollary~32]\label{cor:wtsanti}
Let $\C\subseteq \mat_{n\times m}(\FF_q)$ be a rank-metric code of dimension $\dim(\C)=\ell=k\cdot\max\{m,n\}$. 
The following are equivalent:
\begin{itemize}
\item $\C$ is an optimal anticode,
\item $d_{\ell}(\C)=k$,
\item $d_i(\C)=\lceil i/\max\{m,n\} \rceil$  for $i=1,\ldots,\ell$.
\end{itemize}
\end{cor}

An immediate consequence of Corollary~\ref{cor:wtsMRD} and Corollary~\ref{cor:wtsanti} is the following.

\begin{cor}\label{cor:MRD&optanti}
Let $\C\subseteq \mat_{n\times m}(\FF_q)$ be a rank-metric code. 
Then $\C$ is both MRD and an optimal anticode if and only if $\C=0$ or $\C=\mat_{n\times m}(\FF_q)$.
\end{cor}

A similar result can be obtained for dually quasi-MRD codes. It follows from~\cite[Corollary~18]{DeLaCruz2018} that the dimension of a dually quasi-MRD code $\C\subseteq \mat_{n\times m}(\FF_q)$ is not divisible by $\max\{m,n\}$. Therefore, in the next result we make this assumption without loss of generality.

\begin{cor}[\cite{DeLaCruz2018}, Theorem~22]
Let $\C\subseteq \mat_{n\times m}(\FF_q)$ be a rank-metric code of dimension $\dim(\C)=\ell=k\cdot\max\{m,n\}+r$, with $k\geq 0$ and $0<r<\max\{m,n\}$. The following are equivalent:
\begin{itemize}
\item $\C$ is dually quasi-MRD,
\item $d_1(\C)=\min\{m,n\}-k$ and $d_{r+1}(\C)=\min\{m,n\}+1-k$.
\end{itemize}

Moreover, if $\C$ is dually quasi-MRD, then its generalized weights are:
$$\begin{array}{l} d_1(\C)=\cdots=d_{r}(\C)=\min\{m,n\}-k,\\
d_{r+1+i\cdot\max\{m,n\}}(\C)=\cdots=d_{r+(i+1)\max\{m,n\}}(\C)=\min\{m,n\}+1+i-k\;\;\mbox{ for } i=0,\ldots,k-2,\\
d_{r+1+(k-1)\max\{m,n\}}(\C)=\cdots=d_{\ell}(\C)=\min\{m,n\}.
\end{array}$$
\end{cor}

We already observed that, while there is no easy relation between the minimum distance of $\C$ and $\C^\perp$, the weight distribution of $\C$ determines the weight distribution of $\C^\perp$. The next result shows that the generalized weights of $\C$ determine the generalized weights of $\C^\perp$.

\begin{thm}[\cite{Ravagnani2016u}, Corollary~38]
Let $\C\subseteq\mat_{n\times m}(\FF_q)$ be a rank-metric code of dimension $\dim(\C)=\ell$.
Define the sets $$W_i(\C)=\{d_{i+j\cdot\max\{m,n\}}(\C)\mid j\geq 0,\ 1\leq i+j\cdot\max\{m,n\}\leq \ell\}$$ and 
$$\overline{W}_i(\C)=\{\min\{m,n\}+1-d_{i+j\cdot\max\{m,n\}}(\C)\mid j\geq 0,\ 1\leq i+j\cdot\max\{m,n\}\leq \ell\}.$$
Then
$$W_i(\C^\perp)=\{1,\ldots,\min\{m,n\}\}\setminus\overline{W}_{i+\ell}(\C)$$ for $i=1,\ldots,\max\{m,n\}$.
\end{thm}

\section{$q$-polymatroids and code invariants}\label{qpolym}

$q$-polymatroids are the $q$-analog of polymatroids. In this section we associate to every rank-metric code a $q$-polymatroid for $m\neq n$ and a pair of $q$-polymatroids for $m=n$. We then discuss how several invariants and structural properties of codes, such as generalized weights, the property of being MRD or an optimal anticode, and duality, are captured by the associated $q$-polymatroids. The material of this section is contained in~\cite{Shiromoto2018,Gorla2018p}, but the presentation we give differs at times from the original papers.

We start by giving the definition of a $q$-matroid, the $q$-analog of a matroid.

\begin{defn}[\cite{Jurrius2016}, Definition~2.1]\label{defn:qmatr}
A \textbf{$q$-matroid} is a pair $P=(\FF_q^\ell,\rho)$ where $\rho$ is a function from the set of all subspaces of $\FF_q^\ell$ to $\ZZ$ such that, for all $U,V\subseteq \FF_q^\ell$:
\begin{itemize} \setlength\itemsep{0em}
\item[(P1)] $0\leq \rho(V)\leq\dim(V)$,
\item[(P2)] if $U\subseteq V$, then $\rho(U)\leq \rho(V)$,
\item[(P3)] $\rho(U+V)+\rho(U\cap V)\leq \rho(U)+\rho(V)$.
\end{itemize}
\end{defn}

$q$-polymatroids were defined independently by Shiromoto in~\cite{Shiromoto2018} and by Gorla, Jurrius, Lopez, and Ravagnani in~\cite{Gorla2018p}. The two definitions are essentially equivalent. Here we follow the approach of~\cite{Gorla2018p}.

\begin{defn}[\cite{Gorla2018p}, Definition~4.1]\label{defn:qpoly}
A \textbf{$q$-polymatroid} is a pair $P=(\FF_q^\ell,\rho)$ where $\rho$ is a function from the set of all subspaces of $\FF_q^\ell$ to $\mathbb{R}$ such that, for all $U,V\subseteq \FF_q^\ell$:
\begin{itemize} \setlength\itemsep{0em}
\item[(P1)] $0\leq \rho(V)\leq\dim(V)$,
\item[(P2)] if $U\subseteq V$, then $\rho(U)\leq \rho(V)$,
\item[(P3)] $\rho(U+V)+\rho(U\cap V)\leq \rho(U)+\rho(V)$.
\end{itemize}
\end{defn}

Definition~\ref{defn:qpoly} is a direct $q$-analogue of the definition of an ordinary polymatroid, with the extra property that $\rho(V)\leq \dim(V)$ for all $V\subseteq\FF_q^\ell$. As in the ordinary case, a $q$-matroid is a $q$-polymatroid.

\begin{rmk}\label{rmk:defnShiro}
Definition~\ref{defn:qpoly} is slightly different from the definition of $(q,r)$-polymatroid given by Shiromoto in~\cite[Definition 2]{Shiromoto2018}. However, a $(q,r)$-polymatroid $(E,\rho)$ as defined by Shiromoto corresponds to the $q$-polymatroid $(E,\rho/r)$ according to Definition~\ref{defn:qpoly}. Moreover, a $q$-polymatroid whose rank function takes values in $\mathbb{Q}$ corresponds to a $(q,r)$-polymatroid as defined by Shiromoto up to multiplying the rank function by an $r$ which clears denominators.
\end{rmk}

We now give two simple examples of $q$-matroids.

\begin{ex}
The pair $(\FF_q^\ell,\dim(\cdot))$ is a $q$-matroid, where $\dim(\cdot)$ denotes the function that associates to a vector space its dimension.
\end{ex}

\begin{ex}
For a fixed $U\subseteq\FF_q^\ell$, let $$\rho_U(V)=\dim(V)-\dim(V\cap U^\perp)$$ for $V\subseteq\FF_q^\ell$.
The pair $(\FF_q^\ell,\rho_U)$ is a $q$-matroid.
\end{ex}

One has the following natural notion of equivalence for $q$-polymatroids.

\begin{defn}[\cite{Gorla2018p}, Definition~4.4]\label{defequipolym}
Let $(\FF_q^\ell,\rho_1)$ and $(\FF_q^\ell,\rho_2)$ be $q$-polymatroids. We say that $(\FF_q^\ell,\rho_1)$ and $(\FF_q^\ell,\rho_2)$ are \textbf{equivalent} if there exists an $\FF_q$-linear isomorphism $\varphi:\FF_q^\ell\to \FF_q^\ell$ such that $\rho_1(V)=\rho_2(\varphi(V))$ for all $V\subseteq \FF_q^\ell$. In this case we write $(\FF_q^\ell,\rho_1) \sim (\FF_q^\ell,\rho_2)$.
\end{defn}

The following is the natural notion of duality for $q$-polymatroids.

\begin{defn}[\cite{Gorla2018p}, Definition~4.5]
Let $P=(\FF_q^\ell,\rho)$ be a $q$-polymatroid. For all subspaces $V\subseteq \FF_q^\ell$ define
$$\rho^*(V)=\dim(V)-\rho(\FF_q^\ell)+\rho(V^\perp),$$
where $V^\perp$ is the dual of $V$ with respect to the standard inner product on $\FF_q^\ell$.
We call $P^*=(\FF_q^\ell,\rho^*)$ the \textbf{dual} of the $q$-polymatroid $P$.
\end{defn}

It is easy to show that $P^*$ is indeed a $q$-polymatroid.
The dual of a $q$-polymatroid satisfies the usual properties for a dual. Moreover, duality is compatible with equivalence.

\begin{thm}[\cite{Gorla2018p}, Theorem~4.6 and Proposition~4.7]
Let $P,Q$ be $q$-polymatroids. Then:
\begin{itemize}
\item $P^*$ is a $q$-polymatroid.
\item $P^{**}=P$.
\item If $P \sim Q$, then $P^* \sim Q^*$. 
\end{itemize}
\end{thm}

One can associate $q$-polymatroids to rank-metric codes as follows. In~\cite[Theorem~5.4]{Gorla2018p} it is shown that these are indeed $q$-polymatroids according to Definition~\ref{defn:qpoly}.

\begin{defn}[\cite{Gorla2018p}, Notation~5.3]\label{defn:qpolycode}
Let $\C\subseteq\mat_{n\times m}(\FF_q)$ be a rank-metric code, and let $V\subseteq \FF_q^{\min\{m,n\}}$. 
Define $$\rho_\C(V)=\frac{1}{\max\{m,n\}}(\dim(\C) - \dim(\C(V^\perp))\in\QQ.$$
If $m\neq n$, we associate to $\C$ the $q$-polymatroid $P(\C)=\left(\FF_q^{\min\{m,n\}},\rho_\C\right)$.\\
If $m=n$, we associate to $\C$ the pair of $q$-polymatroids $P(\C)=(\FF_q^n,\rho_\C)$, $P(\C^\textsf{T})=(\FF_q^n,\rho_{\C^\textsf{T}})$.
\end{defn}

Notice that this is slightly different from what is done in~\cite{Gorla2018p}, where a pair of $q$-polymatroids is associated to each rank-metric code. In this chapter, we choose to present the material of~\cite{Gorla2018p} differently, in order to stress the following facts (stated following the notation~\cite[Notation~5.3]{Gorla2018p}): 
\begin{itemize}
\item for $n<m$ the $q$-polymatroid that contains all the relevant information on $\C$ is $(\FF_q^n,\rho_\cc(\C,\cdot))$, 
\item for $n>m$ the $q$-polymatroid that contains all the relevant information on $\C$ is $(\FF_q^m,\rho_\rr(\C,\cdot))$, 
\item for $n=m$ one needs to consider both $(\FF_q^n,\rho_\cc(\C,\cdot))$ and $(\FF_q^n,\rho_\rr(\C,\cdot))$, at least if one wishes to have the property that equivalent codes have equivalent associated $q$-polymatroids.
\end{itemize}

\begin{rmk}
In~\cite[Proposition~3]{Shiromoto2018}, Shiromoto associates a $(q,m)$-polymatroid to any rank-metric code with $n\leq m$. If $n<m$, his definition is equivalent to Definition~\ref{defn:qpolycode}, given what we observed in Remark~\ref{rmk:defnShiro}. For $n=m$, Shiromoto's definition is not equivalent to Definition~\ref{defn:qpolycode}; in particular it is not equivalence-invariant (while Definition~\ref{defn:qpolycode} is). Notice moreover that the original definition by Shiromoto does not contain the assumption that $n\leq m$. However, this hypothesis is used implicitly throughout his paper. Whenever stating the results from~\cite{Shiromoto2018}, we always add the assumption $n\leq m$.
\end{rmk}

The code of~\cite[Example~2.10]{Gorla2018p} shows that the definition of an associated $(q,n)$-polymatroid given by Shiromoto for a rank-metric code $\C\subseteq\mat_{n\times n}(\FF_q)$ is not equivalence-invariant.

\begin{ex}
Let $\C\subseteq\mat_{2\times 2}(\FF_2)$ be the code 
$$\C:=\left\{ \begin{pmatrix} a & a \\ b & b \end{pmatrix}  \mid  a,b \in \FF_2 \right\}.$$
Let $(\FF_2^2,\rho_1)$ and $(\FF_2^2,\rho_2)$ be the $(q,2)$-polymatroids associated to $\C$ and $\C^\textsf{T}$ respectively, according to~\cite[Proposition~3]{Shiromoto2018}. By definition, for any $V\subseteq\FF_2^2$
$$\rho_1(V)=2-\dim(\C(V^\perp))=\dim(V)$$ and 
$$\rho_2(V)=2-\dim(\C^{\textsf{T}}(V^\perp))=\left\{\begin{array}{cl}
\dim(V) & \mbox{if } V=0,\langle(1,1)\rangle,\FF_2^2,\\
2 & \mbox{if } V=\langle(1,0)\rangle,\langle(0,1)\rangle.
\end{array}\right.$$
The natural notion of equivalence for $(q,r)$-polymatroids is the following: $(\FF_q^\ell,\rho_1)$ and $(\FF_q^\ell,\rho_2)$ are equivalent if there exists an $\FF_q$-linear isomorphism $\varphi:\FF_q^\ell\to \FF_q^\ell$ such that $\rho_1(V)=\rho_2(\varphi(V))$ for all $V\subseteq \FF_q^\ell$. Clearly $(\FF_2^2,\rho_1)$ and $(\FF_2^2,\rho_2)$ are not equivalent with respect to such a notion of equivalence. 
\end{ex}

The interest in associating $q$-polymatroids to rank-metric codes comes from the fact that many invariants of rank-metric codes can be computed from the associated $q$-polymatroids. In fact, one could think of (equivalence classes of) $q$-polymatroids as invariants of the rank-metric codes to which they are associated, since equivalent codes are associated to equivalent $q$-polymatroids.

\begin{prop}[\cite{Gorla2018p}, Proposition~6.8]
Let $\C,\D\subseteq \mat_{n\times m}(\FF_q)$ be rank-metric codes. Assume that $\C \sim \D$. If $m\neq n$, then 
$P(\C)\sim P(\D)$. If $n=m$, then one of the following holds:
\begin{itemize}
\item $P(\C) \sim P(\D)$ and $P(\C^\textsf{T}) \sim P(\D^\textsf{T})$,
\item $P(\C) \sim P(\D^\textsf{T})$ and $P(\C^\textsf{T}) \sim P(\D)$.
\end{itemize}
\end{prop}

One can also show that the $q$-polymatroid(s) associated to the rank-metric code $\Gamma(C)$ associated to a vector rank-metric code $C\subseteq\FF_{q^m}^n$ do not depend on the choice of the basis $\Gamma$.

\begin{prop}[\cite{Jurrius2016}, Corollary~4.7 and \cite{Gorla2018p}, Proposition~6.10]
Let $C\subseteq\FF_{q^m}^n$ be a vector rank-metric code, and let $\Gamma,\Gamma'$ be $\FF_q$-bases of $\FF_{q^m}$. Then $$P(\Gamma(C))\sim P(\Gamma'(C)).$$ 
\end{prop}

In the rest of this section, we discuss how to recover various invariants of rank-metric codes from the associated $q$-polymatroids. We start with the simplest invariants, namely the dimension and the minimum distance.  

\begin{prop}[\cite{Gorla2018p}, Proposition~6.1 and Corollary~6.3]
Let $\C\subseteq\mat_{n\times m}(\FF_q)$ be a rank-metric code. Then 
$$\dim(\C)=\max\{m,n\}\cdot\rho_\C\left(\FF_q^{\min\{m,n\}}\right)$$ and  
$$d_{\min}(\C)=\min\{m,n\}+1-\delta$$ where $$\delta=\min\left\{k\left| \rho_\C(V)= \frac{\dim(\C)}{\max\{m,n\}}\mbox{ for all }V \subseteq \FF_q^{\min\{m,n\}}\mbox{ with }\dim(V)=k\right.\right\}.$$
\end{prop}

The next results shows how one can compute the generalized weights of a rank-metric codes from its associated $q$-polymatroid(s). 

\begin{thm}[\cite{Gorla2018p}, Theorem~7.1]
Let $\C\subseteq\mat_{n\times m}(\FF_q)$ be a rank-metric code. Let 
$$\begin{aligned}
d_i(P(\C))=\min\{m,n\}-\max\{\dim(V) \mid V \subseteq \FF_q^{\min\{m,n\}}, \ \dim(\C)-\max\{m,n\}\cdot\rho_{\C}(V) \ge i\},\\ i=1,\ldots,\dim(\C).\end{aligned}$$
If $n\neq m$, then 
$$d_i(\C)=d_i(P(\C))\;\;\mbox{for } i=1,\ldots,\dim(\C).$$
If $n=m$, then $$d_i(\C)=\min\{d_i(P(\C)), \ d_i(P(\C^\textsf{T}))\}\;\;\mbox{for } i=1,\ldots,\dim(\C).$$
\end{thm}

The associated $q$-polymatroid(s) also determine the weight distribution of a rank-metric code. The result is stated in terms of the weight enumerator of the code. 

\begin{defn}
Let $\C\subseteq\mat_{n\times m}(\FF_q)$ be a rank-metric code. The {\bf weight enumerator} of $\C$ is the polynomial $$W_{\C}(x,y)=\sum_{i=0}^{\min\{m,n\}} A_i(\C)x^{i}y^{\min\{m,n\}-i}.$$
\end{defn}

The next theorem was proved by Shiromoto. 

\begin{thm}[\cite{Shiromoto2018}, Theorem~14]
Let $\C\subseteq\mat_{n\times m}(\FF_q)$ be a rank-metric code of dimension $\dim(\C)=\ell$. 
Assume that $n\leq m$. Then
$$W_{\C}(x,y)=x^{n-\ell/m}\left(\sum_{V\subseteq\FF_q^n}(q^mx)^{\rho_\C(\FF_q^n)-\rho_\C(V)} x^{-(\dim(V)-\rho_\C(V))}\right)\prod_{i=0}^{\dim(V)-1}(y-q^ix).$$
\end{thm}

Finally, we state two results that show that the property of being MRD or an optimal anticode can be characterized in terms of the associated $q$-polymatroid(s).

\begin{thm}[\cite{Gorla2018p}, Theorem~6.4 and Corollary~6.6]
Let $\C\subseteq\mat_{n\times m}(\FF_q)$ be a rank-metric code with minimum distance $d_{\min}(\C)=d$.  The following are equivalent:
\begin{itemize}
\item $\C$ is MRD,
\item $\rho_\C(V)=\dim(V)$ for all $V \subseteq \FF_q^{\min\{m,n\}}$ with $\dim(V) \le \min\{m,n\}-d+1$,
\item $\rho_\C(V)=\dim(V)$ for some $V \subseteq \FF_q^{\min\{m,n\}}$ with $\dim(V) = \min\{m,n\}-d+1$.
\end{itemize}
In particular, if $\C$ is MRD with minimum distance $d_{\min}(\C)=d$, then $P(\C)=\left(\FF_q^{\min\{m,n\}},\rho_\C\right)$ where
$$\rho_\C(V)= \left\{ \begin{array}{ll} \min\{m,n\}-d+1 & \mbox{ if } \dim(V) \ge \min\{m,n\}-d+1, \\ 
\dim(V) & \mbox{ if } \dim(V) \le \min\{m,n\}-d+1.\end{array}\right.$$
\end{thm}

The corresponding result for optimal anticodes is the following. Notice that the $q$-polymatroids associated to MRD codes or optimal anticodes are in fact $q$-matroids. 

\begin{thm}[\cite{Gorla2018p}, Theorem~7.2 and Corollary~6.6]
Let $\C\subseteq\mat_{n\times m}(\FF_q)$ be a rank-metric code with $r=\max\rk(\C)$. 
The following are equivalent:
\begin{itemize}
\item $\C$ is an optimal anticode,
\item $\left\{\rho_\C(V)\mid V\subseteq \FF_q^{\min\{m,n\}}\right\}=\left\{0,1,\ldots,r\right\}$, or $\left\{\rho_{\C^\textsf{T}}(V)\mid V \subseteq \FF_q^n\right\}=\left\{0,1,\ldots,r\right\}$ and $m=n$,
\item $\rho_\C\left(\FF_q^{\min\{m,n\}}\right)=r$, or $\rho_{\C^\textsf{T}}(\FF_q^n)=r$ and $m=n$.
\end{itemize}
In particular, if $\C$ is an optimal anticode with $r=\max\rk(\C)$, let 
$$\rho(V)=\dim(V+\langle e_1,\ldots,e_{\min\{m,n\}-r}\rangle)-(\min\{m,n\}-r),$$ 
where $e_i$ denotes the $i$-th vector of the standard basis of $\FF_q^{\min\{m,n\}}$. 

If $m\neq n$, then $P(\C)\sim (\FF_q^{\min\{m,n\}},\rho)$.

If $m=n$, then either $P(\C)\sim(\FF_q^n,\rho)$ or $P(\C^\textsf{T})\sim(\FF_q^n,\rho)$.
\end{thm}

We conclude with a result on associated $q$-polymatroids and duality. The theorem as we state it was proved by Gorla, Jurrius, L\'opez, and Ravagnani in~\cite[Theorem~8.1 and Corollary~8.2]{Gorla2018p}. Shiromoto also proved in~\cite[Proposition~11]{Shiromoto2018} that $P(\mathcal{C})^*=P(\mathcal{C}^\perp)$ for a rank-metric code $\C\subseteq\mat_{n\times m}(\FF_q)$ with $n\leq m$.

\begin{thm}
Let $\C\subseteq\mat_{n\times m}(\FF_q)$ be a rank-metric code and let $C\subseteq\FF_{q^m}^n$ be a vector rank-metric code. Let $\Gamma$ be a basis of $\FF_{q^m}$ over $\FF_q$. Then 
$$P(\mathcal{C})^*=P(\mathcal{C}^\perp)\;\;\;\mbox{and}\;\;\; P(\Gamma(C))^*\sim P(\Gamma(C^\perp)).$$ 
\end{thm}

\bibliographystyle{amsplain}

\end{document}